\documentclass[11pt,reqno]{amsart}

\usepackage{amsmath}
\usepackage{amssymb}
\usepackage{graphicx}
\usepackage{accents}
\usepackage{pstricks,pst-node,pst-coil,pst-plot}
\usepackage{mathtools}
\usepackage{ulem}
\usepackage{color}
\usepackage{dcolumn}
\usepackage{mathrsfs}
\usepackage{slashed}
\usepackage{epstopdf}
\usepackage{bm}
\usepackage{slashed}
\usepackage{dcolumn}

\numberwithin{equation}{section}

\newtheorem{thm}{Theorem}[section]
\newcommand{\be}{\begin{equation}}
\newcommand{\ee}{\end{equation}}
\newcommand{\bee}{\begin{equation*}}
\newcommand{\eee}{\end{equation*}}
\def\m{\mathfrak{m}}

\newtheorem{cla}{Claim}[section]
\newtheorem{prop}{Proposition}[section]
\newtheorem{coro}{Corollary}[section]

\def\d{\mathrm{d}}

\let\<\langle
\let\>\rangle

\newcommand{\definedas}{\mathrel{\raise.095ex\hbox{\rm :}\mkern-5.2mu=}}

\begin{document}


\title[From bending of light to positive mass: a non-PDE perspective]{From bending of light to positive mass:\\ a non-PDE perspective}
\author{Xiaokai He}
\address[Xiaokai He]{School of Mathematics and Computational Science, Hunan First Normal University, Changsha 410205, China}
\email{sjyhexiaokai@hnfnu.edu.cn}
\author{Xiaoning Wu}
\address[Xiaoning Wu]{Institute of Mathematics, Academy of Mathematics and Systems Science and Hua Loo-Keng Key Laboratory, Chinese Academy of Sciences, Beijing 100190, China}
\email{wuxn@amss.ac.cn}
\author{Naqing Xie}
\address[Naqing Xie]{School of Mathematical Sciences, Fudan
University, Shanghai 200433, China}
\email{nqxie@fudan.edu.cn}
\begin{abstract}
Penrose et al. investigated the physical incoherence of the spacetime with negative mass via the bending of light.  Precise estimates of time-delay of null geodesics were needed and played a pivotal role in their proof. In this paper, we construct an intermediate diagonal metric and make a reduction of this problem to a causality comparison in the compactified spacetimes regarding timelike connectedness near the conformal infinities. This different approach allows us to avoid encountering the difficulties and subtle issues Penrose et al. met. It provides a new, substantially simple, and physically natural non-PDE viewpoint to understand the positive mass theorem. This elementary argument modestly applies to asymptotically flat solutions which are vacuum and stationary near infinity.
\end{abstract}

%




\maketitle



\section{Introduction}\label{S1}
One of the fundamental results in classical general relativity is the positive mass theorem. The first complete proof was given by Schoen and Yau \cite{S+YI,S+YII} and the theorem was extended to dimensions up to seven by Eichmair, Huang, Lee and Schoen \cite{EHLR16}. Schoen-Yau's proof was heavily based on the analysis of elliptic partial differential equations (PDE) with many minimal surface techniques. Later, Witten provided an alternative proof using spinors \cite{Wi}. His proof was inspired by supergravity and given in easily understood terms. However, the theory of general relativity itself does not invoke supersymmetry. It is somewhat surprising that albeit these proofs are mathematically sophisticated, they do not pertain the Lorentzian nature of spacetimes. In particular, the arguments seem not to have anything to do with the causality structures.

There appears to be a pursuit of more real physical understanding to the positive mass theorem among physicists. Penrose proposed a scheme to show the negative mass spacetimes are physically inconsistent \cite{P80}. The original idea was to derive a contradiction by certain focusing theorems and time-delay of null geodesics. The test for the pure Schwarzschild was given in \cite{P90}. Ashtekar and Penrose examined more closely what it is about the structure of the spatial infinity $i^0$ that the argument establishes \cite{AP90}. The idea was further carried out by constructing a null line as a fastest causal curve via the Vietoris topology by Penrose, Sorkin and Woolgar in \cite{PSW93} and recently Cameron extended the result to higher dimensions \cite{C20}. The time of flight estimate diverges logarithmically in the $(3+1)-$dimension while it tends to zero in higher dimensions. Cameron successfully found an alternative proof by a comparison argument on the physical spacetimes \cite[(4.16)]{C20} to construct a Minkowskian-null curve which is timelike with respect to the physical metric in the presence of negative mass.

Chru\'{s}ciel and Galloway proved a poor man's version of the positive mass theorem for uniformly Schwarzschildean spacetimes by reducing the problem to the Lorentzian splitting theorem \cite{CG04}. There is also a successful variant of the argument using null lines \cite{Ch04}. Negative mass leads to the existence
of a null line in the spacetime, which then is incompatible with energy conditions. The proofs in \cite{CG04,Ch04} do not invoke the `generic condition' \cite[Page 101]{HE73} required in the focusing theorems.

In this paper, we implement Penrose's original idea to further understand the physical incoherence of the negative mass. The main difficulties in \cite{P80,P90,PSW93,C20} were to obtain the precise estimates of time-delay or retardation of null geodesics. The explicit calculation could be done for spacetimes with symmetry but it is not so easy to have controls for perturbations. Instead of looking into the behavior of null geodesics, we attempt to avoid these subtle issues by considering the timelike connectedness in a neighborhood of the conformal infinity.

Our ambitions in this paper are quite modest. We consider the class of asymptotically flat spacetimes which are vacuum and stationary near infinity. And we assume that the geodesic completeness and the energy conditions in Borde's focusing theorem \cite{Bo87} are satisfied. The main contribution of this paper is to provide a simple reduction of the problem to a causality comparison theorem. This is highly motivated by the very recent work of Cameron and Dunajski on the Penrose property in \cite{CD20}. Firstly, we construct an intermediate metric $\tilde{h}$ with the same mass which is diagonal in the polar coordinates. As the usual procedure \cite[Appendix]{DS90}, one compactifies both the intermediate metric $\tilde{h}$ and the physical metric $\tilde{g}$ using the same conformal factor. Secondly, we make an comparison argument in the conformal spacetime between the compactified metric $h$ and the compactified Minkowski metric $\eta$, but not in the physical spacetime as in \cite{PSW93,C20}. The `generic condition' is still imposed since we need to use certain focusing theorem making our proof quite elementary.

The organization of the manuscript is as follows. In Section \ref{S2}, we give a quick review of Penrose's idea to exclude the negative mass. Section \ref{S3} deals with the reduction of the problem to a causality comparison. And we show that the spacetimes discussed here indeed satisfy this causality comparison which fulfills the contradiction argument by Penrose. Summary and discussion are provided in the last section. We will use geometric units with $c=G=1$ and the spacetime signature convention is assumed to be `mostly minus' $(+,-,-,-)$ as in \cite{P80}.

\section{Weakly asymptotically empty and simple spacetimes}\label{S2}
Suppose that a connected chronological physical spacetime $(\tilde{M},\tilde{g})$ can be conformally included into a strongly causal spacetime with boundary $(M,g)$ such that $\tilde{M}$ is the interior of $M$. Assume that there exists a function $\Omega$ on $M$ such that
\bee
\mbox{(i)}\ \Omega >0 \ \mbox{and}\ g=\Omega^2 \tilde{g} \ \mbox{on}\ \tilde{M} \eee
and
\bee
\mbox{(ii)}\ \Omega=0 \ \mbox{and}\ \d \Omega \neq 0 \ \mbox{along}\ \partial M.\eee

The boundary $\mathscr{I}$ is assumed to consist of two components - two null hypersurfaces, $\mathscr{I}^+$ and $\mathscr{I}^-$, known as the future and the past null infinities, respectively.

A spacetime is said to be \textit{asymptotically simple} \cite[Page 222]{HE73} if, in addition,\\
 \indent (iii) Every null geodesic in $(\tilde{M},\tilde{g})$ is mapped to a null geodesic in $(M,g)$
 which has two endpoints on $\partial M$.

Moreover, if the Ricci tensor of $\tilde{g}$ vanishes on an open neighbourhood of $\partial M$, then the spacetime $(\tilde{M},\tilde{g})$ is said to be  \textit{asymptotically empty and simple} \cite[Page 222]{HE73}. Asymptotically empty and simple spacetimes include the Minkowski spacetime but do not include the Schwarzschild or the Kerr solutions. This suggests that one should define a spacetime to be \textit{weakly asymptotically empty and simple} \cite[Page 225]{HE73} if there is an open set isometric to an open neighbourhood of the boundary of an asymptotically empty and simple spacetime.

Before we state the main result, let us refer to the following sets:
\bee
\begin{split}
I^+(p)&=\{q \in M | \ \mbox{$\exists$ a smooth future directed}\\
&\ \ \ \ \ \ \ \mbox{timelike curve from $p$ to $q$}\},\\
I^-(p)&=\{q \in M | \ p \in I^+(q)\},\\
J^+(p)&=\{q \in M | \ \mbox{$\exists$ a smooth future directed }\\
&\ \ \ \ \ \ \ \mbox{causal curve from $p$ to $q$}\},\\
J^-(p)&=\{q \in M | \ p \in J^+(q)\}.
\end{split}
\eee
We also require that $\mathscr{D}\cup \mathscr{I}$ be globally hyperbolic, where $\mathscr{D}$ denotes
the domain of outer communications $\mathscr{D}=I^-(\mathscr{I}^+) \cap I^+(\mathscr{I}^-)$.

Recall that the metric of the Schwarzschild spacetime with mass $\m$ reads
\be
\tilde{g}_{\m}=\big(\frac{1-\frac{\m}{2r}}{1+\frac{\m}{2r}}\big)^2\d t^2-\big(1+\frac{\m}{2r} \big)^4 \big(\d r^2+r^2\d \theta^2 +r^2\sin^2\theta \d \varphi^2\big).\ee
We say a metric $\tilde{g}$ on $\mathbb{R} \times (\mathbb{R}^3\setminus B)$ is \textit{asymptotically Schwarzschildean} with nonzero mass $\m$ \cite{CG04,C20} if
\be
\tilde{g}-\tilde{g}_{\m}=o\big(|\m| r^{-1}\big)\ee
where $B$ denotes a sufficiently large Euclidean ball centered at the origin.

One expects
\begin{thm}
Suppose that $(\tilde{M},\tilde{g})$ is a weakly asymptotically empty and simple spacetime which satisfies the conditions of Borde's focusing theorem\footnote{The standard energy condition on usual geodesic focusing is a pointwise inequality. This pointwise condition has been weakened by Borde to the requirement that certain integral comes out to be nonnegative \cite{Bo87}. It is not clear yet whether splitting theorems still work in these extended situations.} and is such that $\mathscr{D}\cup \mathscr{I}$ is globally hyperbolic. If the physical metric $\tilde{g}$ is asymptotically Schwarzschildean with mass $\m$, then  $\m$ cannot be negative.
\end{thm}

 Below let us outline Penrose's idea of the proof in \cite{P90}. Assume that the mass $\m$ at $i^0$ were strictly negative.  The key point is to show that \begin{cla}\label{cla1}
The spacetime $(M,g)$ with $\m<0$ does not have the Penrose property\footnote{According to \cite[Definition 2.4]{CD20}, a spacetime $(M,g)$ is called to have the Penrose
property if any points $p \in \mathscr{I}^-$ and $q \in \mathscr{I}^+$ can be timelike connected.}, i.e., there exists a pair of points $p^\prime \in \mathscr{I}^-$ and $q^\ast \in \mathscr{I}^+$ such that $q^\ast \notin I^+(p^\prime)$.\end{cla}
Now assume that the above claim is proved. One can easily find a timelike curve starting from $p^\prime$ and ending at $q^{\ast\ast}$ on $\mathscr{I}^+$, i.e., $q^{\ast\ast} \in I^+(p^\prime) \cap \mathscr{I}^+$. Then there must be a boundary point $\check{q} \in \partial I^+(p^\prime) \cap \mathscr{I}^+$. Since $\mathscr{D}\cup \mathscr{I}$ is globally hyperbolic, there must be a null geodesic from $p^\prime$ to $\check{q}$ lying entirely on $\partial I^+(p^\prime)$  \cite{P72}. On the other hand, according to certain focusing theorems \cite{HP70,Bo87}, there should be a pair of conjugate points between $p^\prime$ and $\check{q}$, and hence there exists a timelike curve from $p^\prime$ and $\check{q}$. This contradicts with $\check{q} \in \partial I^+(p^\prime)$ since $I^+(p^\prime)$ is an open set.

\section{Conformal extensions of stationary spacetimes}\label{S3}
Let $\tilde{g}$ be an asymptotically flat, stationary solution to the vacuum Einstein field equations. We assume that the mass does not vanish and there exists a coordinate chart $(t,x^i)$ near infinity in which the physical metric has the following expansion \cite{BS80,DS90}
\be
\begin{split}\tilde{g}&=\big(1-\frac{2\m}{r}+\frac{2\m^2}{r^2}
+\frac{F}{r^3}\big)\d t^2+\big(4\epsilon_{ijk}\frac{S^jx^k}{r^3}
+\frac{F_i}{r^3}\big)\d t\d x^i\\
&\ -\big(\delta_{ij}(1+\frac{2\m}{r}+
\frac{3}{2}\frac{\m^2}{r^2})+\frac{F_{ij}}{r^3}\big)\d x^i
\d x^j.\\
\end{split}
\ee
Here the constants $\m$ and $S^i$ are the asymptotic mass and the angular momentum respectively, $\epsilon_{ijk}$ is the alternating Levi-Civita symbol, and $r=\sqrt{(x^1)^2+(x^2)^2+(x^3)^2}$ is the Euclidean radius. The functions $F$, $F_i$, $F_{ij}$ into which we have collected the higher order terms are bounded \cite{DS90,D01}.

The stationary metric $\tilde{g}$ admits an analytic conformal extension through null infinity \cite[Theorem 2.8]{D01}. In order to prove our main result, here we write down the details of the construction of the conformal compactification.

Firstly, we construct an intermediate metric $\tilde{h}$ with the same mass which is diagonal in terms of the standard polar coordinates.

\begin{prop}\label{p1}
Assume that $\m \neq 0$. Then there exist positive constants $G>0$ and $R_0>0$ such that, for $r>R_0$, by identifying the physical coordinates $(t,x^i)$, one has the following comparison between metrics in the physical spacetime near infinity:
\be\label{h1}
\tilde{g}\leq \tilde{h}=A(r)\d t^2-B(r)\delta_{ij}\d x^i
\d x^j
\ee
where
\bee
A(r)=1-\frac{2\m}{r}+\frac{G \m^2}{r^2}\eee
and
\bee
B(r)=1+\frac{2\m}{r}+\frac{\m^2}{r^2}.\eee
Therefore, any $\tilde{g}-$causal curve must also be $\tilde{h}-$causal via the identification of the physical coordinates $(t,x^1,x^2,x^3)$ near infinity.\end{prop}
\noindent \textit{Proof:} Suppose $|F|+|F_i|+|F_{ij}|\leq C_1|\m|^3 $ and $|S^j|\leq C_2 \m^2$. Then there exists an $R_0>0$ such that, for $r>R_0$,
\bee
\sum_{i,j}|\frac{F_{ij}}{r^3}\d x^i\d x^j|\leq  \sum_{i,j}\frac{\m^2}{8r^2} \delta_{ij}\d x^i\d x^j, \eee
\bee
\sum_{i}|\frac{F_i}{r^3}\d t \d x^i| \leq  \frac{\m^2}{8r^2} \big(\d t^2+\sum_{i}(\d x^i)^2\big),\eee
and
\bee
|\frac{F}{r^3} \d t^2|  \leq \frac{\m^2}{8r^2} \d t^2.\eee
Moreover,
\bee
\begin{split}
&\ \ \ \ |4\epsilon_{ijk}\frac{S^jx^k}{r^3}\d t \d x^i| \\
&\leq \sum_{i}2|\frac{4C_2\m^2}{r^2}\d t \d x^i|\\
&\leq \m^2\big(\frac{48C_2^2}{\epsilon r^2}\d t^2 +  \frac{\epsilon}{r^2}\sum_{i}(\d x^i)^2\big)\end{split}\eee
for any $\epsilon>0$.

By taking $\epsilon=\frac{1}{4}$, we have
\bee
\begin{split}\tilde{g}&=\big(1-\frac{2\m}{r}+\frac{2\m^2}{r^2}
+\frac{F}{r^3}\big)\d t^2+(4\epsilon_{ijk}\frac{S^jx^k}{r^3}
+\frac{F_i}{r^3})\d t\d x^i\\
&\ \ \ \ -\big(\delta_{ij}(1+\frac{2\m}{r}+
\frac{3}{2}\frac{\m^2}{r^2})+\frac{F_{ij}}{r^3}\big)\d x^i
\d x^j\\
&\leq \Big(1-\frac{2\m}{r}+\frac{2\m^2}{r^2}
+\frac{\m^2}{8r^2}+ \frac{\m^2}{8r^2}+\frac{48\m^2 C_2^2}{\frac{1}{4}r^2}\Big)\d t^2\\
&\ \ \ \ -\big(\delta_{ij}(1+\frac{2\m}{r}+
\frac{3}{2}\frac{\m^2}{r^2}-\frac{\m^2}{4r^2}-\frac{\m^2}{8r^2}-\frac{\m^2}{8r^2})\big)\d x^i
\d x^j\\
&=\big(1-\frac{2\m}{r}+\frac{G \m^2}{r^2})\d t^2 -\big(1+\frac{2\m}{r}+\frac{\m^2}{r^2}\big)\delta_{ij}\d x^i\d x^j.
\end{split}
\eee
Here $G=\frac{9}{4}+192C_2^2$ and this completes the proof. \hfill $\Box$

The metric $\tilde{h}$ can be rewritten as a diagonal form in the polar coordinates $(t,r,\theta,\varphi)$
\be
\label{h2}
\tilde{h}=A\d t^2-B\big(\d r^2 +r^2\d \theta^2+r^2\sin^2\theta \d \varphi^2 \big).\ee

Let
\be\label{to}
r_\ast= \int \sqrt{\frac{B}{A}}\d r=r+2\m\log (\frac{r}{R_0})+O(\frac{1}{r})\ee
where $R_0$ is the positive constant in Proposition \ref{p1}.

We define the retarded and the advanced time coordinates $u$  and $v$ respectively as
\be
u=t-r_\ast, \ \  \ v=t+r_\ast.\ee
and the compactified time and radial coordinates $T$ and $\chi$ are further defined as
\be
T=\arctan v+\arctan u,\ \ \
\chi=\arctan v-\arctan u.
\ee
Equivalently, one has
\be
u=\tan(\frac{T-\chi}{2}), \ \ \ v=\tan(\frac{T+\chi}{2})
\ee
and hence
\be
\d u=\frac{1}{2\cos^2\frac{T-\chi}{2}}(\d T-\d\chi),\
\d v=\frac{1}{2\cos^2\frac{T+\chi}{2}}(\d T+\d\chi).
\ee

The metric $\tilde{h}$ now becomes
\be
\begin{split}
\tilde{h}&=A\d t^2-B\big(\d r^2 +r^2\d \theta^2+r^2\sin^2\theta \d \varphi^2 \big)\\
&=A\d t^2-B\frac{A}{B}\d r_\ast^2   - B\big(r^2\d \theta^2+r^2\sin^2\theta \d \varphi^2 \big)\\
&=A\d u \d v - B\big(r^2\d \theta^2+r^2\sin^2\theta \d \varphi^2 \big)\\
&=\frac{A}{4\cos^2\frac{T+\chi}{2}\cos^2\frac{T-\chi}{2}} \big(\d T^2-\d \chi^2\big)\\
&\ \  -Br^2\big(\d \theta^2+\sin^2\theta \d \varphi^2 \big)\\
&=\Omega^{-2}\Big( \d T^2-\d \chi^2-Br^2\Omega^2\big(\d \theta^2 +\sin^2\theta \d \varphi^2 \big)  \Big)  .\end{split}\ee
where the conformal factor
\be\label{conf}
\Omega=\frac{2\cos\frac{T+\chi}{2}\cos\frac{T-\chi}{2}}{\sqrt{A}}.
\ee

Denote the conformal metric by
\be\label{cm}
\begin{split}
h&=\Omega^2\tilde{h}\\
&=\d T^2-\d \chi^2-Br^2\Omega^2\big(\d \theta^2 + \sin^2\theta \d \varphi^2 \big)\\
&=\d T^2-\d \chi^2-\frac{Br^2}{Ar_*^2}\sin^2\chi\big(\d \theta^2 + \sin^2\theta \d \varphi^2 \big) .
\end{split}
\ee

One compactifies the physical metric $\tilde{g}$ into $g=\Omega^2 \tilde{g}$ using the same conformal factor $\Omega$ as in \eqref{conf} since $\tilde{g}$ is a perturbation of $\tilde{h}$ with the same mass \cite{DS90}. The boundary set $\{\Omega=0\}$ becomes a null hypersurface. It turns out that the identification via the compactified coordinates $(T,\chi,\theta,\varphi)$ keeps the points at null infinities fixed between $\mathscr{I}_g^{\pm}$ and $\mathscr{I}_h^{\pm}$. And we have the following causality comparison result near the conformal infinity.
\begin{coro}\label{c1}
At any point slightly away from the conformal infinity, any $g-$causal curve must also be $h-$causal via the identification of the compactified coordinates $(T,\chi,\theta,\varphi)$ near infinity.\end{coro}
\noindent \textit{Proof:} Note that we have used the same conformal factor $\Omega$ as in \eqref{conf} to compactify the metrics $\tilde{g}$ and $\tilde{h}$. The same $(T,\chi,\theta,\varphi)$ means the same $(t,r,\theta,\varphi)$ for points away from the conformal infinity. Then the proof immediately follows from Proposition \ref{p1}.\hfill $\Box$

Secondly, we make a comparison in the conformal spacetime between the compactified metric $h$ and the compactified Minkowski metric $\eta$. The conformal metric $h$ \eqref{cm} differs from the compactified Minkowski metric $\eta$ \cite[(3.7)]{CD20} only by the function
$\frac{Br^2}{Ar_*^2}$ which appears in front of the angular terms. This function is identically $1$ for Minkowski. In the presence of negative mass, there is a bijection between curves in the $h-$spacetime and curves in the compactified Minkowski spacetime near infinity arising from identifying the compactified coordinates $(T,\chi,\theta,\varphi)$.
The following comparison proposition plays a key role in the argument whose prototype for the negative mass Schwarzschild spacetime can be found in \cite[Proposition 4.3]{CD20}.

\begin{prop}\label{p2}
 Assume that $\m<0$. In a neighborhood of the conformal infinity, an $h-$timelike curve is also Minkowskian-timelike under the identification via the compactified coordinates $(T,\chi,\theta,\varphi)$.
\end{prop}
\noindent \textit{Proof:}
We consider the large $r$ expansion of the function $\frac{B}{A}\frac{r^2}{r^2_\ast}$.
\be
\begin{split}
&\ \ \frac{B}{A}\frac{r^2}{r^2_\ast}\\
&=\big(1+\frac{4\m}{r}+O(\frac{1}{r^2})\big)\Big(1+\frac{2\m\log (\frac{r}{R_0})}{r}+O(\frac{1}{r^2})\Big)^{-2}\\
&=\big(1+\frac{4\m}{r}+O(\frac{1}{r^2})\big)\Big(1+\frac{-4\m\log(\frac{r}{R_0}) }{r}+O\big((\frac{\log(\frac{r}{R_0})}{r^2})^2\big)\Big)\\
&=1+\frac{4\m}{r}+\frac{-4\m\log(\frac{r}{R_0})}{r} +O\big((\frac{\log(\frac{r}{R_0})}{r^2})^2\big)\\
&>1.
\end{split}\ee
This means the conformally metric $h$ can be bounded above by the conformally compactified Minkowski metric $\eta$, i.e., $h<\eta$ for large $r$.
\hfill $\Box$

Now we are in the position to prove Claim \ref{cla1} for asymptotically flat solutions which are vacuum and stationary near infinity.\\

\noindent \textit{Proof of Claim \ref{cla1}.} We use a contradiction argument. If Claim \ref{cla1} were wrong, then for any pair of $p \in \mathscr{I}^-$ and $q \in \mathscr{I}^+$, $p$ and $q$ could be connected by a $g-$timelike curve $C$. One can draw the causal diamond between $p$ and $q$ with respect to the metric $\eta$. It is easy to see that the entire region will be covered by the coordinates $(T,\chi,\theta,\varphi)$ if $p$ and $q$ are close enough to $i^0$. With the aid of the causality comparison results in Corollary \ref{c1} and Proposition \ref{p2}, it is clear that the $g-$causal diamond between $p$ and $q$ is covered by the $\eta-$causal diamond, i.e., the possibility of any $g$-timelike curve $C(p,q)$ running out off the neighborhood of $i^0$ we consider has been ruled out. And consequently, the $g-$timelike curve $C$ is also $h-$timelike and further $\eta-$timelike via the identification of the compactified coordinates $(T,\chi,\theta,\varphi)$.

\begin{figure}[htp]
\begin{center}
\includegraphics[width=1.0\textwidth]{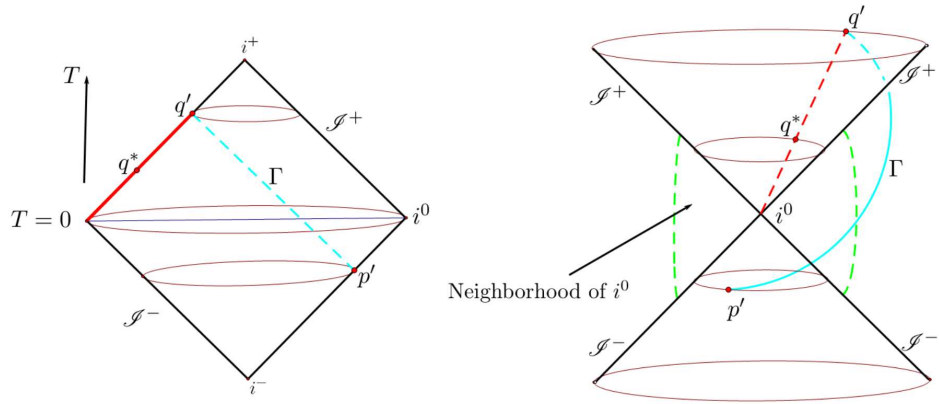}
\caption{\label{Penrose1} Compactified Minkowski spacetime. The point $q^\ast$ exactly lies near $i^0$ and to the past of $q^\prime$ and on the same null generator of $\mathscr{I}^+$ (i.e., the red interval) where the antipodal point $q^\prime$ is the unique future endpoints of null geodesics from $p^\prime$ which enter the Minkowski spacetime. Here $\Gamma$ denotes a null geodesic between the antipodal points $p^\prime$ and $q^\prime$ entering the Minkowski spacetime.}
\end{center}
\end{figure}

On the other hand, this leads to a contradiction. The points $p^\prime=(-\varepsilon,\pi-\varepsilon,\frac{\pi}{2},0) $ and $q^\ast=(\varepsilon,\pi-\varepsilon,\frac{\pi}{2},\pi)$ cannot be $\eta-$timelike connected in the compactified Minkowski spacetime according to Proposition 3.3 in \cite{CD20}. In fact, $q^\ast$ exactly lies near $i^0$ and to the past of $q^\prime$ and on the same null generator of $\mathscr{I}^+$ (i.e., the red interval) where the antipodal point $q^\prime$ is the unique future endpoints of null geodesics from $p^\prime$ which enter the Minkowski spacetime, see Figure \ref{Penrose1} below (also cf. Figure 7 in \cite[Page 10]{CD20}).

This completes the proof of Claim \ref{cla1} and therefore excludes the negative mass for physically realistic non-singular spacetimes.\hfill $ \Box$

\section{Summary and Discussion}
Following the idea of Penrose \cite{P90}, we study the physical incoherence of negative mass for spacetimes.
The spirit of the argument here is very much similar to the work in \cite{PSW93,C20}, both relying on the properties of the spacetime in a neighborhood of the conformal infinity, whereas the methods of Schoen-Yau and Witten impose conditions on certain initial data set. The authors of \cite{PSW93,C20} made use of a causality comparison in the physical spacetime (cf. (4.16) in \cite{C20}) and they must take more care about the convergence when constructing the null line. The difference is that here we make the reduction of this problem to causality comparison on the compactified spacetime which avoids us meeting the difficulties to precisely estimate the time-delay or deflection angle along null geodesics. The constructed intermediate metric is diagonal in the polar coordinates making the comparison simple. Albeit the argument here is restricted to the class of asymptotically flat solutions which are vacuum and stationary near infinity, this provides a physically natural and thus somewhat easily accessible to physicists elementary point of view to understand the positive mass theorem. Such spacetimes are weakly asymptotically empty and simple \cite[Page 225]{HE73}. It is not clear at all whether there exist physically reasonable non-stationary solutions fully satisfying all the conditions imposed in \cite{PSW93,C20,Ch04}. The extension to more general metrics dropping the stationary assumption by using the current argument still awaits a more complete and delicate justification. Clearly it is beyond the scope of the current manuscript.

There is a drawback in this work. We have no mention at all to the zero mass case. Unlike the proofs in \cite{CG04,Ch04}, the `generic condition' plays an essential role here by using the focusing theorem. However, the rigidity of the positive mass theorem says the zero mass forcing the spacetime being Minkowski in which the `generic condition' fails. Therefore, our approach is not able to prove anything regarding the zero mass case.

\section*{Acknowledgments}
 X. He was partially supported by the Natural Science Foundation of Hunan Province (Grant 2018JJ2073). X. Wu was partially supported by the National Natural Science Foundation of China (Grants 11731001 and 11575286). N. Xie was partially supported by the National Natural Science Foundation of China (Grant 11671089).



\begin{thebibliography}{10}
\bibitem{S+YI} R.~Schoen, S.~-T.~Yau, On the proof of the positive mass conjecture in general relativity,  Commun. Math. Phys. 65 (1979) 45-76.
\bibitem{S+YII} R.~Schoen, S.~-T.~Yau, Proof of the positive mass theorem. II,  Commun. Math. Phys. 79 (1981)  231-260.
\bibitem{EHLR16} M.~Eichmair, L.~-H. Huang, D.~Lee, R.~Schoen, The spacetime positive mass theorem in dimensions
less than eight, J. Eur. Math. Soc. 18 (2016), 83-121.
\bibitem{Wi} E.~Witten, A new proof of the positive energy theorem, Commun. Math. Phys. 80 (1981) 381-402.
\bibitem{P80} R.~Penrose, On Schwarzschild causality - a problem for `Lorentz covariant' general relativity, in: F.~J.~Tipler (Ed.), Essays in General Relativity: A. Taub Festschrift, Academic, New York, 1980, pp. 1-12.
\bibitem{P90} R.~Penrose, Light ray near $i^0$: a new mass-positivity theorem, Twistor Newsletter 30 (1990) 1-5.
\bibitem{AP90} A.~Ashtekar, R.~Penrose, Mass positivity from focussing and the structure of $i^0$, Twistor Newsletter 31 (1990) 1-5.
\bibitem{PSW93} R.~Penrose, R.~Sorkin, E.~Woolgar, A positive mass theorem based on the focusing and retardation of
null geodesics, arXiv:9301015 [gr-qc].
\bibitem{C20} P.~Cameron, Positivity of mass in higher dimensions, arXiv:2010.05086 [gr-qc].
\bibitem{CG04} P.~Chru\'{s}ciel, G.~Galloway, A poor man's positive energy theorem, Class. Quantum Grav. 21 (2004) L59-L63.
\bibitem{Ch04} P.~Chru\'{s}ciel, A poor man's positive energy theorem: II. Null geodesics, Class. Quantum Grav. 21 (2004) 4399-4415.
\bibitem{HE73} S.~W.~Hawking, G.~F.~Ellis, The Large Scale Structure of Space-time, Cambridge University Press, Cambridge, 1973.
\bibitem{Bo87} A.~Borde, Geodesic focusing, energy conditions and singularities, Class. Quantum Grav. 4 (1987) 343-356.
\bibitem{CD20} P.~Cameron, M.~Dunajski, On Schwarzschild causality in higher dimensions, Class. Quantum Grav. 37 (2020) 225002.
\bibitem{DS90} T.~Damour, B.~Schmidt, Reliability of perturbation theory in general relativity, J. Math. Phys. 31 (1990) 2441-2453.
\bibitem{P72} R.~Penrose, Techniques of Differential Topology
in General Relativity, SIAM, Philadelphia, 1972.
\bibitem{HP70} S.~W.~Hawking, R.~Penrose, The singularities gravitational collapse and cosmology, Proc. Roy. Soc. Lond. A 314  (1970) 529-548.
\bibitem{BS80} R.~Beig, W.~Simon, The stationary gravitational field near
spatial infinity, Gen. Rel. Grav. 12 (1980) 1003-1013.
\bibitem{D01} S.~Dain, Initial data for stationary spacetimes near spacelike infinity, Class. Quantum Grav. 18 (2001) 4329-4338.








\end{thebibliography}
\end{document}